\documentclass[aps,prd,twocolumn,amsmath,amssymb,amsfonts,nofootinbib,superscriptaddress,altaffilletter]{revtex4-1}
\usepackage{graphicx}
\usepackage{hyperref}
\hypersetup{
  bookmarksopen=true
}
\usepackage{amssymb}
\usepackage{amsmath}
\usepackage{color}
\usepackage{supertabular}
\usepackage{dcolumn}

\def\sci#1#2{#1\times10^{#2}}

\def\RAJ{\textrm{RA}_{\textrm J2000}}
\def\DECJ{\textrm{DEC}_{\textrm J2000}}

\begin{document}

\title{ 
Results from an extended Falcon all-sky survey for continuous gravitational waves
}

\author{Vladimir Dergachev}
\email{vladimir.dergachev@aei.mpg.de}
\affiliation{Max Planck Institute for Gravitational Physics (Albert Einstein Institute), Callinstrasse 38, 30167 Hannover, Germany}
\affiliation{Leibniz Universit\"at Hannover, D-30167 Hannover, Germany}

\author{Maria Alessandra Papa}
\email{maria.alessandra.papa@aei.mpg.de}
\affiliation{Max Planck Institute for Gravitational Physics (Albert Einstein Institute), Callinstrasse 38, 30167 Hannover, Germany}
\affiliation{Leibniz Universit\"at Hannover, D-30167 Hannover, Germany}
\affiliation{University of Wisconsin Milwaukee, 3135 N Maryland Ave, Milwaukee, WI 53211, USA}

\begin{abstract}
We present the results of an all-sky search for continuous gravitational wave signals with frequencies in the 200-600\,Hz range and frequency derivative (spindown) from $\sci{-1}{-8}$ through $\sci{1.11}{-9}$\,Hz/s. Together with the results from \cite{allsky3}, this search completes the all-sky survey for frequencies between 20 to 600\,Hz on O1 data. It also demonstrates the scalability of our search on a parameter space 26 times larger than previously considered. The results presented here complement the LIGO O2 data results \cite{lvc_O2_allsky,Palomba:2019vxe} with comparable when not better sensitivity and do not rely on data with irregularities in the noise-subtraction procedure. 
We establish strict upper limits which hold for worst-case signal parameters and dedicated upper limits for generic $\approx 0$ spindown signals, such as those expected from boson condensates around black holes.
\end{abstract}

\maketitle

\begin{figure*}[htbp]
\begin{center}
  \includegraphics[width=7in]{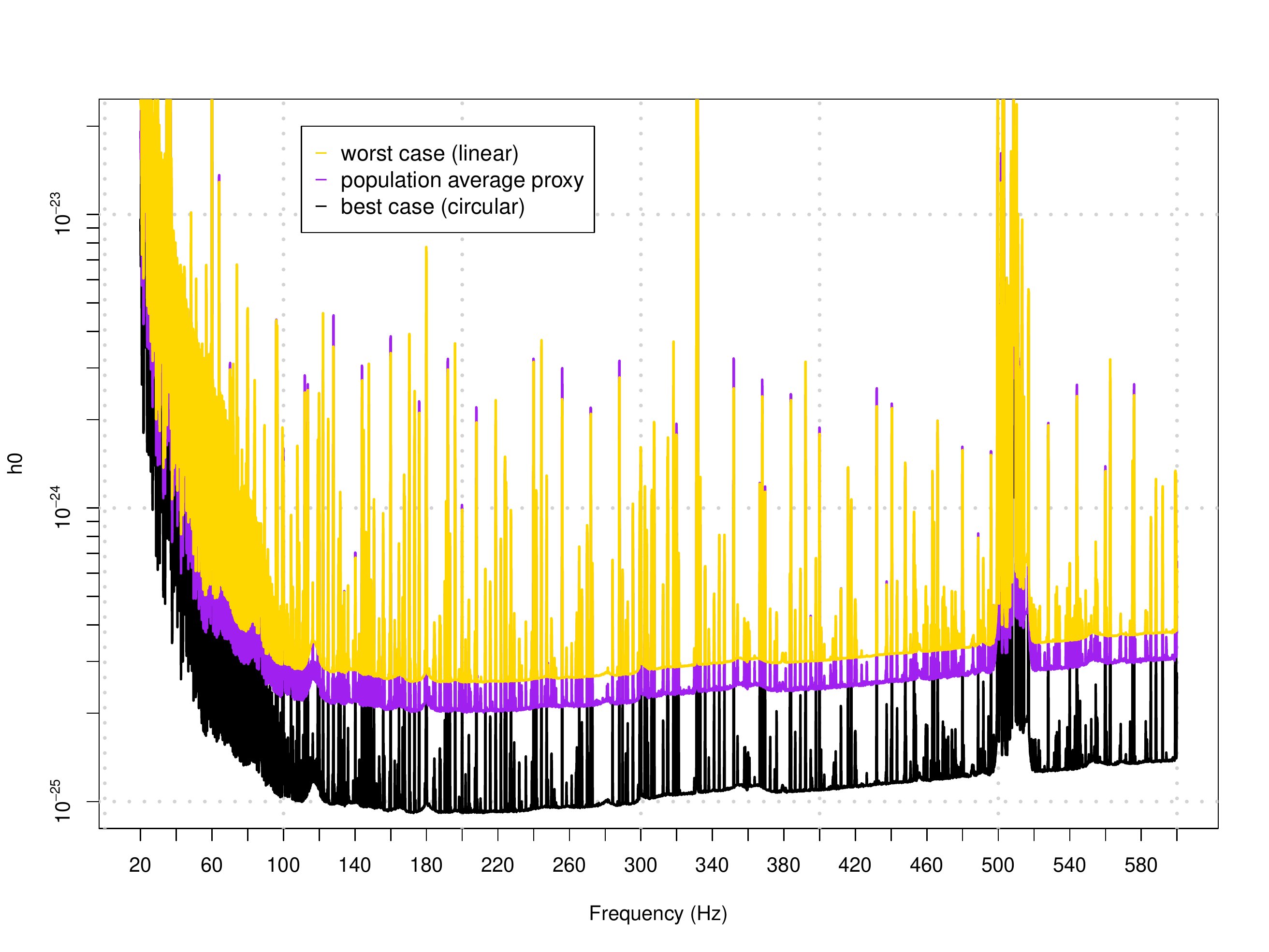}
 \caption{O1 upper limits. The dimensionless strain (vertical axis) is plotted against signal frequency. Looking at the right side of the plot, the upper (yellow) curve shows worst-case upper limits, the next lower curve (purple) shows the population average proxy, followed by the black curve showing best-case upper limits (circularly polarized signals). The worst-case and best-case upper limits are maximized over sky and all intrinsic signal parameters for each frequency band displayed. For completeness we include Falcon results in the frequency range 20-200 Hz from \cite{allsky3}.}
\label{fig:O1_upper_limits}
\end{center}
\end{figure*}

Broad-band all-sky searches for continuous gravitational waves are computationally challenging. The most sensitive searches rely on clever search methods, computationally efficient algorithms and much computing power \cite{allsky3, EHO1, Walsh:2019nmr, Walsh:2016hyc, depths}. 

The loosely coherent approach that we have adopted for our broadest and fast-turnaround surveys has proven to be very successful with short coherence time lengths ($\approx$ half an hour)  \cite{loosely_coherent, loosely_coherent2, loosely_coherent3}. Its recent extension to longer coherent timescales, the Falcon search,  represents a breakthrough in performance and sensitivity \cite{allsky3}. 

The first Falcon search \cite{allsky3} demonstrates the new method investigating the frequency range 20-200\,Hz, all-sky, on data from LIGO's first observation run (O1) \cite{aligo, o1_data, losc}. Here we extend the search to the frequency region 200-600\,Hz. The frequency derivative in both searches is from $\sci{-1}{-8}$ through $\sci{1.11}{-9}$\,Hz/s. The same five-stage pipeline is used, starting with coherence length of 4 hours. 

Even with the most efficient search strategies, continuous wave searches take a much longer time to complete than searches for transient signals and this time significantly grows with increasing signal frequency. In particular, since the number of sky templates scales quadratically with frequency, the size of the parameter space of this search is 26 times greater than that of \cite{allsky3}. As a consequence it took a longer time to carry out this search, in comparison with \cite{allsky3}. We use O1 data because this search was started before the O2 data \cite{o2_data} had been made public.  This search demonstrates that the efficiency of Falcon is maintained in a significantly larger production run.

The full list of outliers that survive all the stages is available in \cite{data}. Table \ref{tab:Outliers} shows a summary of this list. The summary excludes all outliers within 0.01\,Hz of multiples of 0.5\,Hz, which are induced by 0.25\,Hz combs of instrumental lines \cite{O1LowFreq}. The remaining outliers are summarized by displaying the largest SNR outlier in every 0.1\,Hz band. The top 7 outliers are caused by hardware injected simulated signals whose parameters are listed in Table \ref{tab:injections}. The large majority are due to large hardware artifacts. A few dozen outliers cannot be ascribed to instrumental causes based on $h(t)$ data alone. Studies based on physical and environmental monitoring channels, that are not public, could shed more light on their origin. 

We compute 95\% confidence level upper limits on the intrinsic gravitational wave amplitude $h_0$ of a quasi-monochromatic signal with slow evolution in frequency that can be approximated by a linear model. These are shown in Figure \ref{fig:O1_upper_limits}. The upper limit data is available in computer-readable format in \cite{data}. We also provide upper limits for $\approx 0$ spindown signals, relevant for boson condensates around black holes \cite{boson1,boson2}.
We leave it to the interested reader to constrain from our upper limits physical quantities of interest, based on the specific model they wish to consider.

The upper limits are computed using the universal statistic algorithm \cite{universal_statistics} and are valid in all frequency bands and for the entire sky. The worst case and circular polarization 95\% confidence level upper limits are obtained by maximizing upper limits established for individual sky locations, spindowns and frequency bands. Therefore, they are applicable to any subset of the searched parameter space. 
The population-average proxy upper limits are provided for ease of comparison with other search results  
\cite{O1LowFreq, O1AllSky2, EHO1, Palomba:2019vxe}. They are computed as weighted average of upper limits from individual polarizations. We use the same weighting as in \cite{allsky3}. 

\begin{figure}[htbp]
\includegraphics[width=3.3in]{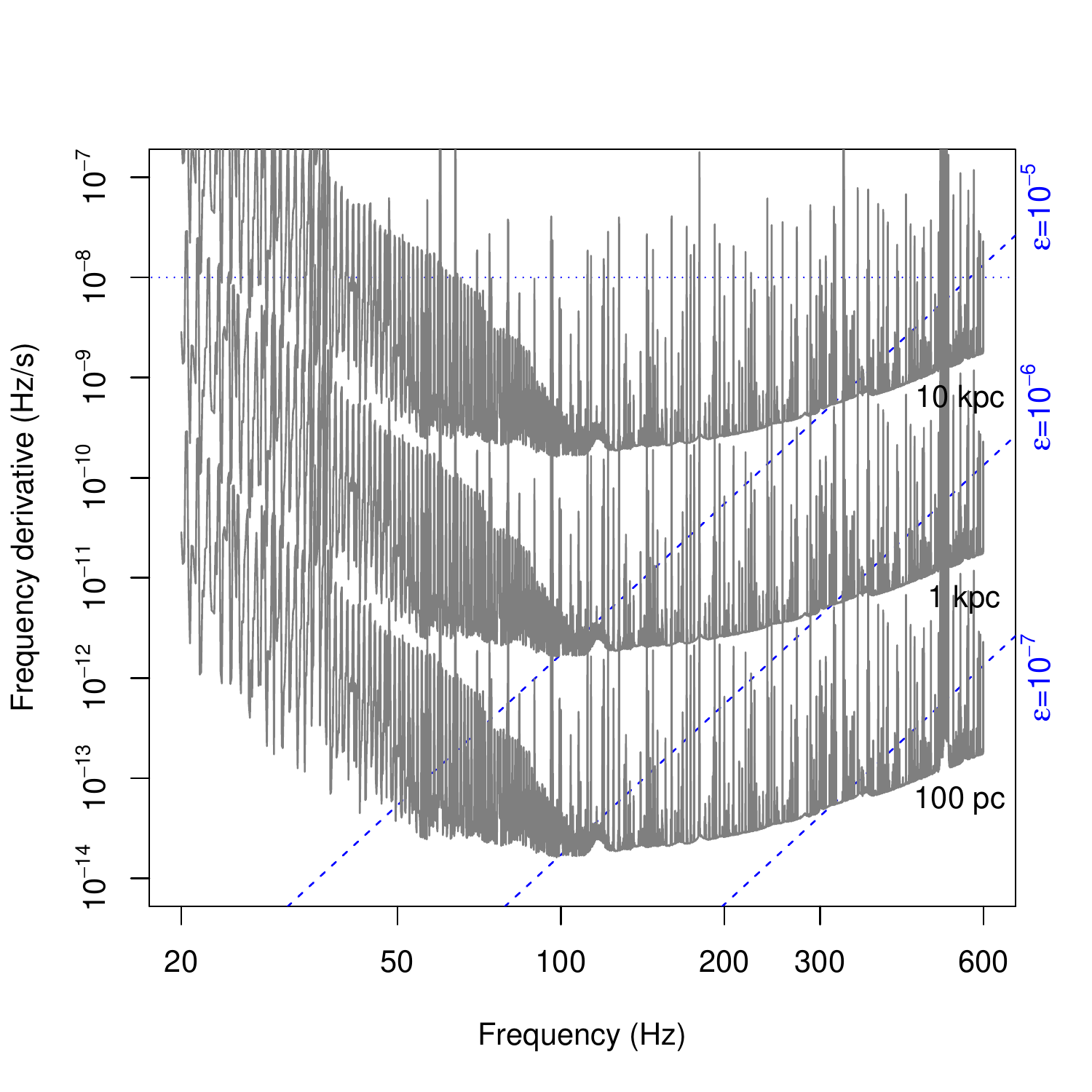}
\caption[Spindown range]{
\label{fig:spindown_range}
Range of the search for neutron stars
spinning down solely due to gravitational radiation.  This is a
superposition of two contour plots.  The grey solid lines are contours of the maximum distance at which a neutron
star could be detected as a function of gravitational wave frequency
$f$ and its derivative $\dot{f}$.  The dashed lines 
are contours of the corresponding ellipticity
$\epsilon(f,\dot{f})$. The fine dotted line marks the maximum spindown searched. Together these quantities tell us the
maximum range of the search in terms of various populations \cite{O1LowFreq, O1AllSky2} }
\end{figure}

Figure \ref{fig:spindown_range} presents circular polarization upper limits converted into maximum distance curves \cite{O1LowFreq, O1AllSky2}. At the high end of the frequency range we are sensitive to a source with $10^{-6}$ equatorial ellipticity up to 2.7\,kpc away. It is known that neutron stars can readily support equatorial ellipticities of more than $10^{-6}$ \cite{crust_limit, crust_limit2}.

Compared to the O2 data-set, the O1 data-set is less sensitive in many frequency bands, has larger instrumental contamination and half the accumulated time. 
Nevertheless, the upper limits presented here are comparable to the LIGO-Virgo collaboration results \cite{lvc_O2_allsky} and to \cite{Palomba:2019vxe} on O2 data. In the 500-600\,Hz frequency range our upper limits are more constraining than those of \cite{lvc_O2_allsky, Palomba:2019vxe}.

The O2 data used in \cite{lvc_O2_allsky, Palomba:2019vxe} was subject to a cleaning procedure  that removed substantial amount of spurious instrumental noise. The cleaning procedure was originally designed for short-lived signals \cite{O2_cleaning} but was subsequently applied to the entire data stream.

The basic idea of the cleaning procedure is to fit $h(t)$ data to many witness data streams and to subtract the polluting contributions. This requires the transfer function from witness data to $h(t)$ to be estimated. The transfer function is non stationary and the estimation is performed in-sample separately on a high number of short time intervals -- and correspondingly small amounts of data. Accidental correlations, which are more likely to happen when the fits are done using small data-sets, can lead to ``over-cleaning", an example of which are the spikes below the noise floor level in Figure 1 of \cite{lvc_O2_allsky}. In general this procedure will contribute an additional systematic uncertainty to the calibration. Whereas we have no reason to believe that such uncertainty amounts to more than a few percent, it is not discussed in \cite{lvc_O2_allsky}\footnote{Public O2 data is limited to the cleaned $h(t)$ data. The release of the uncleaned $h(t)$ and the witness channels would allow for an independent assessment of the extent of this effect.}. Our results provide strict upper limits at a comparable level of sensitivity, derived from a data-set not treated with such procedure.


We also note the interesting discussion of signals from ultra-light bosons in \cite{Palomba:2019vxe}. The population average upper limits in \cite{Palomba:2019vxe} are derived from a signal population that is {\it much} larger than that of boson signals. 
It is not immediately clear to us how to translate a population-average upper limit into an upper limit for a very tiny subset of that population.
We use instead a strict worst-case upper limit from the $\approx 0$ spindown search results {\it only}. 


In conclusion, we have applied the Falcon pipeline to O1 open data. We explore frequencies from 20 to 600\,Hz and frequency derivative from $\sci{-1}{-8}$ through $\sci{1.11}{-9}$\,Hz/s, not relying on data with irregularities in the noise-subtraction procedure. We establish strict upper limits over the entire spindown range and  provide dedicated upper limits for generic $\approx 0$ spindown signals, such as those expected from boson condensates around black holes.

{
\begin{table*}[htbp]
\begin{center}
\tiny
\begin{tabular}{D{.}{.}{2}D{.}{.}{3}D{.}{.}{5}D{.}{.}{4}D{.}{.}{4}D{.}{.}{4}l}\hline
\multicolumn{1}{c}{Idx} & \multicolumn{1}{c}{SNR}   &  \multicolumn{1}{c}{Frequency} & \multicolumn{1}{c}{Spindown} &  \multicolumn{1}{c}{$\RAJ$}  & \multicolumn{1}{c}{$\DECJ$} & Description \\
\multicolumn{1}{c}{}	&  \multicolumn{1}{c}{}	&  \multicolumn{1}{c}{Hz}	&  \multicolumn{1}{c}{nHz/s} & \multicolumn{1}{c}{degrees} & \multicolumn{1}{c}{degrees} & \\
\hline \hline
1 & 990 & 575.16353 & -0.002 & 215.266 & 3.420 & Hardware injection ip2\\
2 & 870 & 52.80832 & 0.006 & 306.634 & -83.997 & Hardware injection ip5\\
3 & 637 & 191.03126 & -8.652 & 351.425 & -33.552 & Hardware injection ip8\\
5 & 577 & 265.57554 & -0.006 & 71.522 & -56.250 & Hardware injection ip0\\
6 & 387 & 146.16934 & -6.710 & 359.608 & -65.199 & Hardware injection ip6\\
7 & 376 & 38.47793 & -6.235 & 332.323 & -14.679 & Hardware injection ip12\\
9 & 300 & 108.85718 & -0.006 & 178.641 & -33.400 & Hardware injection ip3\\
21 & 76 & 440.01500 & -3.465 & 64.741 & -89.269 & Coincident bin-centered lines in H1 and L1 at 440 Hz\\
27 & 54 & 99.97667 & -5.115 & 100.314 & -41.321 & coincident contamination in LHO and LLO\\
28 & 50 & 31.51238 & -5.619 & 226.702 & -23.180 & Heavy contamination, 0.25 Hz comb in H1, 31.512 Hz line in L1\\
35 & 39 & 568.01983 & -4.590 & 10.582 & -89.728 & Coincident bin-centered lines at 568 Hz in H1 and L1\\
40 & 27 & 65.51035 & -5.419 & 198.120 & -40.763 & 0.25 Hz comb of instrumental lines\\
43 & 25 & 460.31067 & -6.248 & 248.229 & 25.004 & Bin-centered line in L1 at 460.3 Hz\\
44 & 24 & 213.01333 & -5.794 & 200.927 & -76.281 & Strong bin-centered line in L1 at 213 Hz \\
45 & 24 & 346.79879 & -9.548 & 118.917 & -66.016 & Strong bin-centered line in L1 at 346.8 Hz\\
46 & 23 & 499.28421 & -3.294 & 113.162 & -85.284 &  \\
47 & 23 & 32.69785 & -9.940 & 45.757 & -37.300 & Lack of coherence, contamination in H1\\
48 & 23 & 336.03237 & -9.206 & 46.881 & -64.585 & Strong bin-centered line in H1 at 336 Hz\\
49 & 23 & 489.40904 & -0.619 & 105.644 & -87.021 & \\
51 & 21 & 82.51584 & -3.994 & 157.388 & -46.681 & Lack of coherence, 0.25 Hz comb in H1\\
52 & 21 & 81.52983 & -6.310 & 332.731 & -45.648 & 0.25 Hz comb of instrumental lines\\
53 & 21 & 400.91107 & -7.790 & 236.031 & -14.678 & Bin-centered line in L1 at 400.9 Hz\\
54 & 21 & 107.13643 & -6.677 & 12.094 & -57.316 & coincident artifacts at 107.12 Hz in H1 and L1\\
55 & 21 & 395.10804 & -7.269 & 235.216 & -2.870 & Strong bin-centered line in L1 at 395.1 Hz\\
56 & 21 & 504.04105 & -8.631 & 45.942 & -34.696 & Contaminated spectrum in L1\\
57 & 21 & 494.64592 & -1.252 & 290.051 & -6.967 & Bin-centered line in L1 at 494.6 Hz\\
58 & 21 & 575.21330 & -5.623 & 244.809 & -26.514 & Induced by injected pulsar 2\\
59 & 20 & 389.25710 & -4.085 & 133.784 & 8.790 & Strong bin-centered line in L1 at 389.3 Hz\\
61 & 20 & 450.97950 & -2.673 & 138.341 & -71.593 & Large broad line in H1\\
63 & 20 & 113.01128 & 1.044 & 304.688 & 9.253 & 0.25 Hz comb of instrumental lines\\
66 & 19 & 90.65642 & -6.944 & 302.827 & 59.828 & no coherence, disturbed H1 spectrum\\
67 & 19 & 542.80440 & -2.160 & 225.365 & 73.334 & \\
69 & 19 & 372.38427 & -7.227 & 194.433 & -5.782 & Strong bin-centered line in L1 at 372.4 Hz\\
70 & 19 & 263.70538 & -6.265 & 267.151 & -6.564 & Spectrum disturbance in L1 at 263.68 Hz\\
71 & 19 & 232.01442 & -4.702 & 253.569 & -32.984 & Strong bin-centered line in H1 at 232 Hz\\
72 & 19 & 62.80672 & -7.985 & 276.491 & -12.515 & Sharp bin-centered line in L1 at 62.8 Hz\\
73 & 19 & 45.01809 & -1.227 & 186.589 & 28.604 & Lines in H1 and L1 at 45 Hz, contaminated spectrum\\
74 & 19 & 558.82419 & -3.890 & 96.652 & -42.306 & \\
75 & 19 & 523.31879 & -4.381 & 62.264 & -29.181 & \\
76 & 19 & 507.88723 & -9.340 & 215.354 & -22.646 & Disturbed spectrum in H1, L1, artifact in H1\\
77 & 19 & 500.47631 & -3.069 & 223.181 & 22.283 & Disturbed spectrum\\
78 & 18 & 329.90057 & -5.369 & 209.627 & -43.310 & Strong bin-centered line in L1 at 329.9 Hz\\
79 & 18 & 595.69484 & -7.548 & 135.507 & -40.958 & \\
80 & 18 & 133.30755 & -6.548 & 124.281 & -50.348 & Line in L1 at 133.33 Hz\\
81 & 18 & 431.80112 & -2.415 & 139.466 & -73.854 & Bin-centered line in L1 at 431.8 Hz\\
82 & 18 & 49.96416 & -9.906 & 335.191 & -18.085 & Highly contaminated spectrum\\
83 & 18 & 437.51154 & -8.248 & 34.774 & -81.896 & \\
84 & 18 & 389.32134 & -9.310 & 240.622 & -40.377 & Strong bin-centered line in L1 at 389.3 Hz\\
85 & 18 & 494.79751 & -3.385 & 51.134 & -85.496 & Weak peak in H1 \\
86 & 18 & 164.68348 & -4.927 & 48.010 & -9.672 & Line in L1 at 164.7 Hz\\
87 & 18 & 265.60671 & -6.777 & 55.182 & -46.651 & Induced by injected pulsar 0\\
88 & 18 & 86.51503 & -5.252 & 56.836 & -36.012 & Coincident lines at 86.5 Hz, 0.25 Hz comb, sloping spectrum in L1\\
89 & 18 & 289.83178 & -7.869 & 82.058 & -42.351 & Strong bin-centered line in L1 at 289.8 Hz\\
90 & 18 & 48.98773 & -4.410 & 43.189 & -22.949 & Highly contaminated H1 and L1 spectrum near 49 Hz\\
91 & 18 & 437.47776 & 0.556 & 243.755 & 77.132 & Sharp line in H1 \\
92 & 18 & 493.92431 & -5.498 & 54.876 & -81.477 & \\
93 & 18 & 586.05732 & -8.790 & 106.536 & 56.297 & \\
94 & 18 & 437.16341 & -2.285 & 46.974 & -55.430 & \\
95 & 18 & 474.80102 & 0.452 & 82.433 & -66.225 & Bin-centered line in L1 at 474.8 Hz\\
96 & 18 & 192.83187 & -7.790 & 248.222 & 46.044 & Contaminated H1 spectrum\\
97 & 18 & 520.38580 & -8.852 & 199.175 & -17.384 & Broad peak in H1\\
98 & 18 & 505.25220 & -8.515 & 217.942 & -6.080 & Disturbed spectrum\\
99 & 18 & 583.20390 & -3.969 & 264.475 & 19.116 & \\
101 & 18 & 536.07116 & 0.860 & 340.569 & 32.085 & \\
102 & 18 & 313.91131 & -2.856 & 279.757 & -19.852 & \\
104 & 18 & 263.30291 & 0.552 & 248.540 & -12.794 & \\
105 & 18 & 292.73093 & -8.140 & 91.585 & -71.773 & Strong bin-centered line in L1 at 292.7 Hz\\
107 & 18 & 54.12124 & -9.790 & 225.480 & -16.962 & Sharp bin-centered line in L1 at 54.1 Hz\\
108 & 17 & 402.62921 & -9.381 & 78.985 & -29.987 & \\
109 & 17 & 403.69274 & -5.406 & 267.397 & 49.040 & \\
110 & 17 & 576.71810 & -0.994 & 149.248 & -87.368 & Bin-centered line in L1 at 576.7 Hz\\
111 & 17 & 520.16849 & -5.810 & 139.840 & -38.511 & Bin-centered line in L1 at 520.2 Hz\\
112 & 17 & 307.06911 & -3.319 & 37.475 & -36.676 & \\
114 & 17 & 235.71326 & -5.585 & 248.190 & -9.510 & Strong bin-centered line in L1 at 235.7 Hz\\
115 & 17 & 573.81279 & -8.410 & 245.184 & 20.706 & Bin-centered line in L1 at 573.8 Hz\\
116 & 17 & 360.81832 & -7.048 & 248.604 & -26.722 & Strong bin-centered line in L1 at 360.8 Hz\\
117 & 17 & 569.11624 & -2.552 & 101.200 & -31.105 & \\
118 & 17 & 499.18839 & -5.898 & 112.912 & -4.852 & \\
119 & 17 & 264.88175 & -7.019 & 128.380 & 30.645 & \\
120 & 17 & 545.75305 & -8.185 & 148.278 & -17.831 & Bin-centered line in L1 at 545.8 Hz\\
121 & 17 & 476.19888 & 0.706 & 187.584 & 76.698 & \\
122 & 17 & 527.75761 & -7.119 & 42.136 & -87.901 & \\
123 & 17 & 583.70009 & -4.719 & 279.324 & 21.711 & \\
124 & 17 & 496.30591 & 1.094 & 70.496 & -62.031 &  \\
125 & 17 & 594.72548 & -5.665 & 191.759 & -57.949 & \\
126 & 17 & 91.71446 & -8.506 & 128.301 & -28.413 & Disturbed spectrum in H1 and L1\\
127 & 17 & 233.30599 & -3.310 & 248.211 & 21.634 & Strong bin-centered line in L1 at 233.3 Hz\\
128 & 17 & 383.51603 & -8.977 & 238.020 & -20.427 & Strong bin-centered line in L1 at 383.5 Hz\\
129 & 17 & 519.40183 & -3.510 & 297.494 & -4.499 &  \\
130 & 17 & 338.13859 & -1.535 & 354.725 & -18.423 & Strong bin-centered line in H1 at 338.1 Hz\\
131 & 17 & 572.25638 & -5.327 & 103.114 & -29.292 & \\
132 & 17 & 550.96513 & -7.510 & 177.002 & 41.547 &  \\
133 & 17 & 371.20547 & -10.065 & 80.438 & 40.981 & \\
134 & 17 & 489.90223 & 0.727 & 243.597 & 41.975 & \\
135 & 17 & 249.69852 & 1.056 & 299.094 & 75.656 & Strong bin-centered line in L1 at 249.7 Hz, near ecliptic pole\\
136 & 17 & 489.75017 & -2.048 & 223.733 & 52.532 & \\
137 & 17 & 363.36911 & -1.706 & 223.220 & -25.441 & \\
139 & 17 & 457.94864 & -7.819 & 170.523 & 54.718 & \\
140 & 17 & 409.12979 & -3.598 & 294.263 & 31.772 & Bin-centered line in L1 at 409.1\\
141 & 17 & 340.16488 & -1.373 & 21.686 & 33.143 & \\
142 & 17 & 365.08911 & 0.356 & 268.928 & 32.316 &  \\
143 & 17 & 53.93050 & -2.956 & 212.788 & -24.064 & Disturbed H1 spectrum\\
144 & 17 & 595.42453 & -6.160 & 105.539 & -39.152 & \\
145 & 17 & 498.90070 & 0.860 & 254.355 & 31.267 & \\
146 & 17 & 107.66022 & -9.660 & 24.618 & -6.139 & Sharp line in L1 at 107.7 Hz\\
147 & 17 & 587.52000 & -1.356 & 216.851 & -53.248 & \\
148 & 17 & 264.19402 & -5.990 & 200.539 & -21.531 & Strong bin-centered line in L1 at 264.2 Hz\\
149 & 17 & 520.01878 & -9.752 & 24.734 & -38.491 & \\
150 & 17 & 251.47765 & -8.619 & 84.701 & -25.529 & \\
151 & 17 & 437.63312 & -0.619 & 297.096 & 18.966 & Bin-centered line in L1 at 437.6\\
152 & 17 & 335.05630 & -7.027 & 110.027 & -54.943 & \\
153 & 17 & 517.76355 & -4.727 & 177.760 & 42.038 & Bin-centered line in L1 at 517.8 Hz\\

\hline
\end{tabular}
\caption[Outliers that passed the detection pipeline]{Outliers that passed the detection pipeline excluding outliers within 0.01\,Hz of 0.25 Hz combs of instrumental lines. Only the highest-SNR outlier is shown for each 0.1\,Hz frequency region. Outliers marked with ``line'' have strong narrowband disturbances near the outlier location.
Signal frequencies refer to epoch GPS $1130529362$. For completeness we include Falcon outliers in the frequency range 20-200 Hz from \cite{allsky3}.} 
\label{tab:Outliers}
\end{center}
\end{table*}
}

\begin{table}[htbp]
\begin{center}
\begin{tabular}{lD{.}{.}{6}rD{.}{.}{5}D{.}{.}{4}}
\hline
Label & \multicolumn{1}{c}{Frequency} & \multicolumn{1}{c}{Spindown} & \multicolumn{1}{c}{$\RAJ$} & \multicolumn{1}{c}{$\DECJ$} \\
 & \multicolumn{1}{c}{Hz} & \multicolumn{1}{c}{nHz/s} & \multicolumn{1}{c}{degrees} & \multicolumn{1}{c}{degrees} \\
\hline \hline
ip0   &  265.575533  & $\sci{-4.15}{-3}$   &   71.55193     &  -56.21749 \\
ip2   &  575.163521  & $\sci{-1.37}{-4}$   &  215.25617     &    3.44399 \\
ip3   &  108.857159  & $\sci{-1.46}{-8}$   &  178.37257     &  -33.4366  \\
ip5   & 52.808324  & $\sci{-4.03}{-9}$     &  302.62664     &  -83.83914 \\
ip6   &  146.169370 & $\sci{-6.73}{0\mathrm{\;\;\;}}$     &  358.75095     &  -65.42262 \\
ip8   &  191.031272 & $\sci{-8.65}{0\mathrm{\;\;\;}}$     &  351.38958     &  -33.41852 \\
ip10  &   26.341917 & $\sci{-8.50}{-2}$    &  221.55565     &   42.87730 \\
ip11  &   31.424758 & $\sci{-5.07}{-4}$    &  285.09733     &  -58.27209 \\
ip12  &   38.477939 & $\sci{-6.25}{0\mathrm{\;\;\;}}$     &  331.85267     &  -16.97288 \\
\hline
\end{tabular}
\caption[Parameters of hardware injections]{Parameters of the hardware-injected simulated continuous wave signals during the O1 data run (epoch GPS $1130529362$). Because the interferometer configurations were largely frozen in a preliminary state after the first discovery of gravitational waves from a binary black hole merger, the hardware injections were not applied consistently. There were no injections in the H1 interferometer initially, and the initial injections in the L1 interferometer used an actuation method with significant inaccuracies at high frequencies.}
\label{tab:injections}
\end{center}
\end{table}


The search was performed on the ATLAS cluster at AEI Hannover. We thank Bruce Allen, Carsten Aulbert and Henning Fehrmann for their support. We also thank Heinz-Bernd Eggenstein and Badri Krishnan for helpful comments and suggestions.

This research has made use of data, software and/or web tools obtained from the LIGO Open Science Center (\url{https://losc.ligo.org}), a service of LIGO Laboratory, the LIGO Scientific Collaboration and the Virgo Collaboration. LIGO is funded by the U.S. National Science Foundation. Virgo is funded by the French Centre National de Recherche Scientifique (CNRS), the Italian Istituto Nazionale della Fisica Nucleare (INFN) and the Dutch Nikhef, with contributions by Polish and Hungarian institutes.


\newpage
\begin{thebibliography}{99}

\def\etal{{\it et al.}}

\bibitem{allsky3} 
 Sensitivity Improvements in the Search for Periodic Gravitational Waves Using O1 LIGO DataSensitivity Improvements in the Search for Periodic Gravitational Waves Using O1 LIGO Data,
  V.~Dergachev and M.~A.~Papa,
  Phys.\ Rev.\ Lett.\  {\bf 123}, no. 10, 101101 (2019)
  
\bibitem{Palomba:2019vxe} 
Direct constraints on ultra-light boson mass from searches for continuous gravitational waves,
   C.~Palomba {\it et al.},
  arXiv:1909.08854.

\bibitem{lvc_O2_allsky}
All-sky search for continuous gravitational waves from isolated neutron stars using Advanced LIGO O2 data,
B.~P.~Abbott \etal\ (LIGO Scientific Collaboration and Virgo Collaboration), Phys.\ Rev.\ D {\bf 100} 024004 (2019).

\bibitem{EHO1} 
  First low-frequency Einstein@Home all-sky search for continuous gravitational waves in Advanced LIGO data,
  B.~P.~Abbott {\it et al.} (LIGO Scientific and Virgo Collaborations),
  Phys.\ Rev.\ D {\bf 96}, no. 12, 122004 (2017)

\bibitem{Walsh:2019nmr} 
    Optimizing the choice of analysis method for all-sky searches for continuous gravitational waves with Einstein@Home,
    S.~Walsh, K.~Wette, M.~A.~Papa and R.~Prix,
  Phys.\ Rev.\ D {\bf 99}, no. 8, 082004 (2019)

\bibitem{Walsh:2016hyc} 
  Comparison of methods for the detection of gravitational waves from unknown neutron stars,
  S.~Walsh {\it et al.},
  Phys.\ Rev.\ D {\bf 94}, no. 12, 124010 (2016)

\bibitem{depths}  
Fast and Accurate Sensitivity Estimation for Continuous-Gravitational-Wave Searches,
C.~Dreissigacker, R.~Prix, K.~Wette, Phys.\ Rev.\ D {\bf 98}, 084058 (2018)

  
\bibitem{loosely_coherent} 
On blind searches for noise dominated signals: a loosely coherent approach,
V.~Dergachev, Class.\ Quantum Grav.\ {\bf 27}, 205017 (2010).

\bibitem{loosely_coherent2} 
Loosely coherent searches for sets of well-modeled signals,
V.~Dergachev,
arXiv:1807.02351

\bibitem{loosely_coherent3} 
Loosely coherent searches for medium scale coherence lengths,
V.~Dergachev,
Phys.\ Rev.\ D {\bf 85}, 062003 (2012)  


\bibitem{o1_data} LIGO Open Science Center, \url{https://doi.org/10.7935/K57P8W9D}, (2018)


\bibitem{aligo} Advanced LIGO, J.~Aasi \etal\ (LIGO Scientific Collaboration), Class.\ Quantum Grav.\  {\bf 32} 7 (2015) 

\bibitem{losc} M.~Vallisneri \etal\ The LIGO Open Science Center, proceedings of the 10th LISA Symposium, University of Florida, Gainesville, May 18-23, 2014, arxiv:1410.4839

\bibitem{o2_data} LIGO Open Science Center, \url{https://doi.org/10.7935/CA75-FM95}, (2019)

\bibitem{data} See EPAPS Document No. [number will be inserted by
publisher] for numerical values of upper limits. Also at \url{www.aei.mpg.de/continuouswaves/O1Falcon20_600}

\bibitem{O1LowFreq}
All-sky Search for Periodic Gravitational Waves in the O1 LIGO Data, 
B.~P.~Abbott \etal\ (LIGO Scientific Collaboration and Virgo Collaboration),
Phys.\ Rev.\ D {\bf 96}, 062002 (2017)

\bibitem{boson1} Black Hole Superradiance Signatures of Ultralight Vectors,
M.~Baryakhtar,~R. Lasenby,~M.Teo, Phys.\ Rev.\ D {\bf 96}, 035006s

\bibitem{boson2} Black Hole Mergers and the QCD Axion at Advanced LIGO,
A.~Arvanitaki, M.~Baryakhtar, R.~Lasenby, S.~Dimopoulos, S.~Dubovsky, Phys.\ Rev.\ D {\bf 95}, 043001

\bibitem{universal_statistics}
A Novel Universal Statistic for Computing Upper Limits in Ill-behaved Background,
V.~Dergachev, 
Phys.\ Rev.\ D \textbf{87}, 062001 (2013).


\bibitem{O1AllSky2}
Full Band All-sky Search for Periodic Gravitational Waves in the O1 LIGO Data,
B.~P.~Abbott \etal\ (LIGO Scientific Collaboration and Virgo Collaboration), Phys.\ Rev.\ D {\bf 97} 102003 (2018).

\bibitem{crust_limit} 
Breaking Strain of Neutron Star Crust and Gravitational Waves, 
C.~J.~Horowitz and K.~Kadau, Phys.\ Rev.\ Lett.\ \textbf{102}, 191102 (2009).

\bibitem{crust_limit2} 
Maximum elastic deformations of relativistic stars,
N.~K.~Johnson-McDaniel and B.~J.~Owen,
Phys. Rev. D {\bf 88}, 044004 (2013)



\bibitem{O2_cleaning}  J.C.~Driggers \etal\  (LIGO Scientific Collaboration Instrument Science Authors), Improving astrophysical parameter
estimation via offline noise substraction for Advancd LIGO,
Phys.\ Rev.\ D {\bf 99}, 042001 (2019).

\end{thebibliography}
\end{document}